# Self-Assembly of Diamondoid Molecules and Derivatives

## (MD Simulations and DFT Calculations)

## Yong Xue [1] and G.Ali Mansoori [2]

University of Illinois at Chicago, (M/C 063) Chicago, IL 60607-7052, USA.

E-Mails:  (1) *xueyong37@gmail.com*

(2) *mansoori@uic.edu*

**Abstract:** We report self-assembly and phase transition behavior of lower diamondoid molecules and their primary derivatives using molecular dynamic (MD) simulation and density functional theory (DFT) calculations. Two lower diamondoids (adamantane and diamantane), three adamantane derivatives (amantadine, memantine and rimantadine) and two artificial molecules (Adamantane+Na and Diamantane+Na) are studied separately in 125-molecule simulation systems. We performed DFT calculations to optimize their molecular geometries and obtain atomic electronic charges for the corresponding MD simulation, by which we obtained self-assembly structures and simulation trajectories for the seven molecules. Radial distribution functions and structure factors studies showed clear phase transitions for the seven molecules.

**Keywords:** adamantane, DFT, diamondoids, MD simulation, molecular building block, nanotechnology, self-assembly.

# 1. Introduction:

Diamondoid molecules (also known as cage-hydrocarbons) and their derivatives have been recognized as molecular building blocks in nanotechnology [1-10]. They have been drawn more and more researchers' attentions to their highly symmetrical and strain free structures, controllable nanosturctural characteristics, non-toxicity and their applications in producing variety of nanosturctural shapes, in molecular manufacturing, in nanotechnology and in MEMS [6, 8]. It is important and necessary to study self-assembly of these molecules in order to obtain reference data, such as temperature, pressure, structure factor, bonding properties, etc. for application in nanotechnology e.g. building molecular electronic devices.

Two lower diamondoids (adamantane and diamantane), three adamantane derivatives (amantadine, memantine and rimantadine) and two artificial molecules (substituting one hydrogen atom in adamantane and diamantine with one sodium atom) are studied in this report. We classified them into three groups as shown in Table I.

**Table I.** Molecular formulas and structures of Adamantane, Diamantane, Memantine, Rimantadine, Amantadine, Optimized ADM•Na and Optimized DIM•Na molecules. In these figures blacks represent –C, whites represent –H, Blues represent –N and purples represent –Na.

| Group 1 | | Group 2 | | | Group 3 | |
|---|---|---|---|---|---|---|
| Adamantane | Diamantane | Amantadine | Rimantadine | Memantine | Optimized ADM•Na | Optimized DIM•Na |
| $C_{10}H_{16}$ | $C_{14}H_{20}$ | $C_{10}H_{17}N$ | $C_{11}H_{20}N$ | $C_{12}H_{21}N$ | $C_{10}H_{15}Na$ | $C_{14}H_{19}Na$ |
| 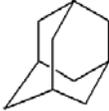 | 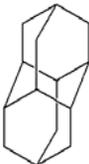 | 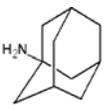 | 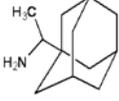 | 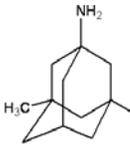 | 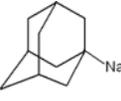 | 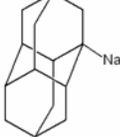 |
| 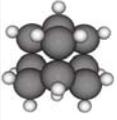 | 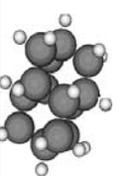 | 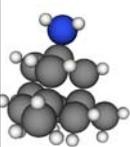 | 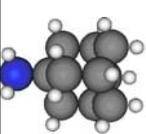 | 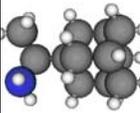 | 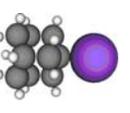 | 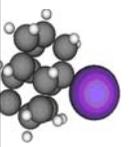 |

Group 1: Adamantane (*ADM*) and Diamantane (*DIM*), the lowest two diamondoids. Due to their six or more linking groups, they have found major applications as templates and as molecular building blocks in nanotechnology, polymer synthesis, drug delivery, drug targeting, DNA-directed assembly, DNA-amino acid nanostructure formation, and host-guest chemistry [1-10]. However these diamondoids do not have good electronic properties which are necessary for building molecular electronics, but some of their derivatives do. Group 2: Memantine, Rimantadine and Amantadine, the three derivatives of adamantane, which have medical applications as antiviral agents and due to their amino groups, they could be treated as molecular semiconductors [8, 9]. Group 3: ADM•Na, DIM•Na, the two artificial molecules, substituting one hydrogen ion in adamantane and diamantane with a sodium ion which could have potential applications in NEMS and MEMS [8-10]. This report is aimed to study the self-assembly and phase transition properties of these seven molecules for further study of their structures and possibly building molecular electronic devices with them.

We performed density functional theory (DFT) calculations to optimize theses seven molecules initial structures and obtained atomic electronic charges [11-16]. Then we performed molecular dynamic (MD) simulation in the study of their self-assembly and phase transition behaviors. MD simulation methods have been broadly used in studying dynamics of molecules, in spite of its classic approximation [17-21]. As typical plastic crystal or say molecular crystal, the phase transition behaviors of adamantane molecules have been studied by other researchers using MD simulation method [22, 23]. Those studies mainly have focused on the transition of the adamantane from FCC to BCC crystal structure or vice versa, and both of these are in the solid phase. Self-assembly, however, is a transition process from random state to an ordered state [1]. In the case of our interest, this process is similar to condensation transition from gas to liquid state, and then freezing transition from liquid state to solid state, which is a border range of phase transitions. For this reason we chose Optimized Potentials for Liquid Simulations All Atom (OPLS-AA) force field in our MD simulations. OPLS-AA force field has shown good results not just in the gas and liquid states but also in crystalline phases in a large range of temperatures [17]; moreover, OPLS-AA is a reliable force field for carbohydrates [18] and, obviously, for hydrocarbons. In the OPLS-AA force field, the non-bonded interactions are represented by the general Lennard-Jones plus Columbic form as shown by Eq. 1, where $E_{ij}$ is the interaction energy of two sites for both intermolecular and intramolecular non-bonded cases.

$$E_{ij} = \sum_{i,j} \left\{ 4\varepsilon_{ij} \left[ \left( \frac{\sigma_{ij}}{r_{ij}} \right)^{12} - \left( \frac{\sigma_{ij}}{r_{ij}} \right)^{6} \right] + \frac{q_i q_j e^2}{r_{ij}} \right\} f_{ij} \tag{1}$$

The combining rules used along with this equation are $\sigma_{ij} = (\sigma_{ii}\sigma_{jj})^{1/2}$ and $\varepsilon_{ij} = (\varepsilon_{ii}\varepsilon_{jj})^{1/2}$. For intermolecular interactions (when $i, j$ corresponds to different molecules) $f_{ij} = 1.0$. While for intramolecular non-bonded interactions between all pairs of atoms ($i < j$) separated by three or more bonds, the 1, 4-interactions are scaled down by $f_{ij} = 0.5$.

Our MD simulations are mainly performed based on the Gromacs package [19-21] and in order to perform simulations involving small molecules, we added to it the necessary ITFs "included topology files" for the seven molecules we studied. Moreover, Gromacs package is capable enough to accept and apply OPLS-AA force field for MD simulations as well as the fact that it is a fast package with flexible characteristics.

Firstly, we report a brief summary; vacuum simulation strategy was used at the beginning to determine how many molecules were suitable for our simulation and obtain equilibrated stable structures. Then we built periodic boundary simulation box; and we used fast PME (Particle-Mesh Ewald) electrostatics method for the preparation of gas state while we chose cut-off technique for electrostatics in the self-assembly/freezing procedure.

To simulate freezing transition, we applied the annealed simulation technique to gradually decease the system temperature. V-rescale temperature coupling was applied throughout and in all the simulations. This is a temperature-coupling method using velocity rescaling with a stochastic term [18]. The van der waals force cut-off and neighbor-searching list distance were, both, set to 2.0 nm, which was large enough for the systems to be isolated from each other under the periodic boundary conditions.

Furthermore, in order to obtain the geometries and atomic charges for the MD simulations of all the seven molecules we applied the DFT calculation which is a widely used method in quantum chemistry and electronic structure theory.

## 2. Density Functional Theory (DFT) Calculations

We performed density functional theory (DFT) calculations to optimize the initial structures of all the seven molecules by using Gaussian 03 package [15] and also we obtained their atomic electronic charges for the MD simulations.

The B3YLP exchange-correlation functional [12] method was chosen with cc-pVDZ basis [14], for all the molecules except for adamantane+Na and diamantane+Na for which 6-311+G(d, p) basis [13] was used since Na is not included in cc-pVDZ basis; NBO (Natural Bond Orbital) analysis [16] was added to calculate the atomic electronic charges which were used as reference to set the atomic charges of nitrogen atom and sodium atom, i.e. -0.76 and 0.65, respectively (in electronic charge units: Coulomb). The atomic charges of carbon and hydrogen atoms were mainly opted from default OPLS-AA force field.

## 3. Molecular Dynamic (MD) Simulation Procedures and Results

The first step was to determine the number of molecules in the simulation box which was suitable to resemble a NVT ensemble. We performed short (20-40 picoseconds) MD simulations of different numbers of adamantane molecules, (i.e. 8, 27, 64, 125, 216, 343, 512, 729 molecules) in vacuum with the temperature set at 100 K, and we successfully obtained all the stable structures as shown in Figure 1.

**Figure 1**. *Snapshots from MD simulations using various numbers of adamantine molecules ($2^3$, $3^3$, $4^3$, $5^3$, $6^3$, $7^3$, $8^3$, $9^3$)*

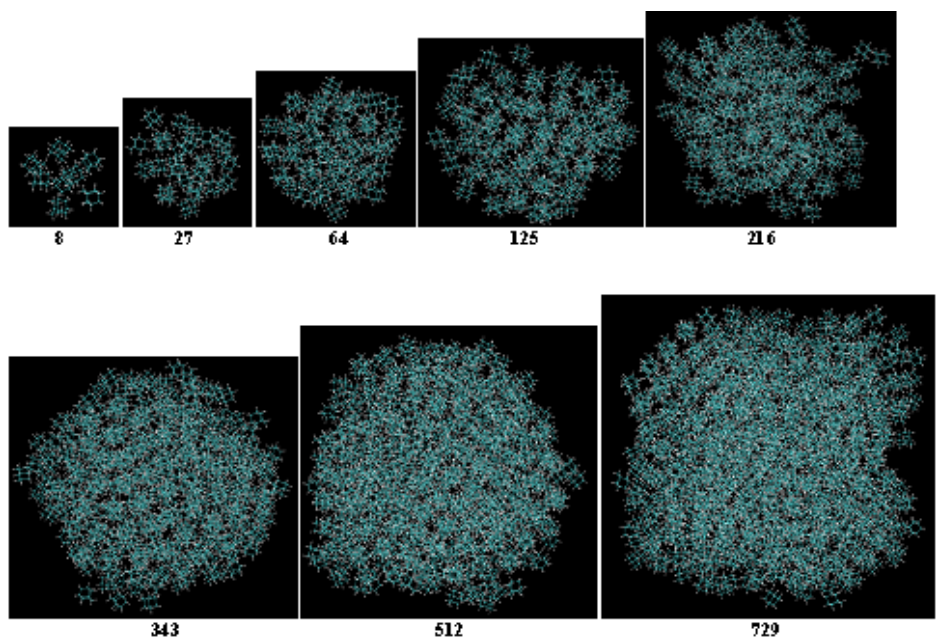

Although those structures are stable, they may not be used for structural analysis, since the simulations are rather too short. And we found that 125-molecule system is large enough to represent the self-assembly behavior, which also can save calculation cost, therefore, we chose 125 molecules in what is reported in the rest of this paper.

From this point on, all the MD simulations involved 125 molecules in the seven different types of diamondoids and their derivatives.

In order to perform self-assembly simulations, we developed the following procedure for the seven types of molecules:

(**a**). A MD simulation box with 5x5x5=125 molecules was built by the intrinsic tools in the Gromacs software as shown in Figure 2-a.

> **Figure 2.** Stages of simulation procedure: (a) Initial 5x5x5 MD simulation box; (b) Molecules relaxed in vacuum at 100 K after a short MD simulation; (c) Boxed simulation system with overall gas density; (d) Gas phase as a result of equilibrating simulation in the NVT ensemble; (e) The liquid state; (f) Final self-assembly of molecules to solid state.

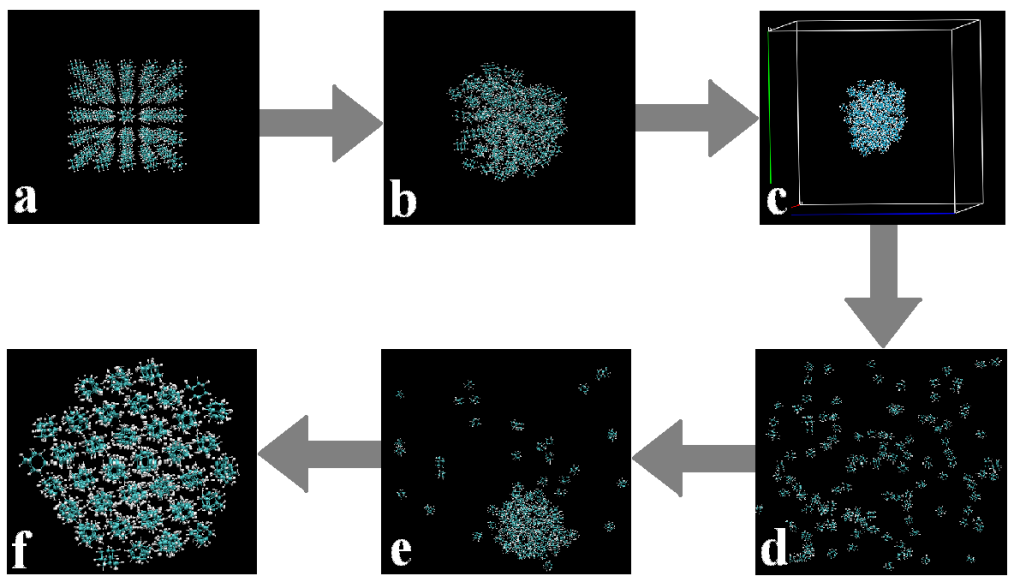

(**b**) Then we performed a short MD simulation by letting all the molecules relax in a vacuum at 100 K (like Figure 1) to obtain the stable self-assembled structure as shown in Figure 2-b.

(**c**) We boxed those stable structures and set the distances from molecules to the boundaries of simulation boxes to about +3 nm. In this manner we could build simulation systems with reasonable densities which are in the range of their gas states and also make the systems isolated from the adjacent simulation boxes under the periodic boundary conditions as shown in Figure 2-c.

(**d**) Longer (more than 1000 picoseconds) equilibrating simulations were performed in the NVT ensembles, and we applied PME method with cut-off at 2.0 nm and high temperatures (in the range of 500 – 700 K) to make sure the entire system went to the gas state, *i.e.* all the molecules separated from each other and distributed randomly in the simulation box as shown in Figure 2-d.

By applying the above four-step procedure we could equilibrate the system in a gas state. As a result, we eliminated the problem due to directly equilibrating the initially prepared molecular system, Figure 2-a, at high temperature. The equilibrated and stable-structure gas state, Figure 2-d, was then ready for cooling down towards the self-assembly.

(**e**) The next step of the simulation was the cooling down the equilibrated and stable-structure gas state, Figure 2-d, from the gas to the liquid state as shown in Figure 2-e.

(**f**) Further cooling down of the system resulted in the complete self-assembly of all the molecules (to solid state) as shown in Figure 2-f.

The snapshots of the gas-liquid-solid MD simulations for 125 molecules (stages d, e and f of Figure 2) of each of the seven molecules (*Adamantane, Diamantane, Amantadine, Rimantadine, Memantine, Adamantane+Na., Diamantane+Na*) are reported in Figure 3.

> **Figure 3**. Self-assembly snapshots of 125 molecules of the seven compounds (from top: Adamantane, Diamantane, Amantadine, Rimantadine, Memantine, Adamantane+Na., Diamantane+Na) as temperature [K] is decreased.

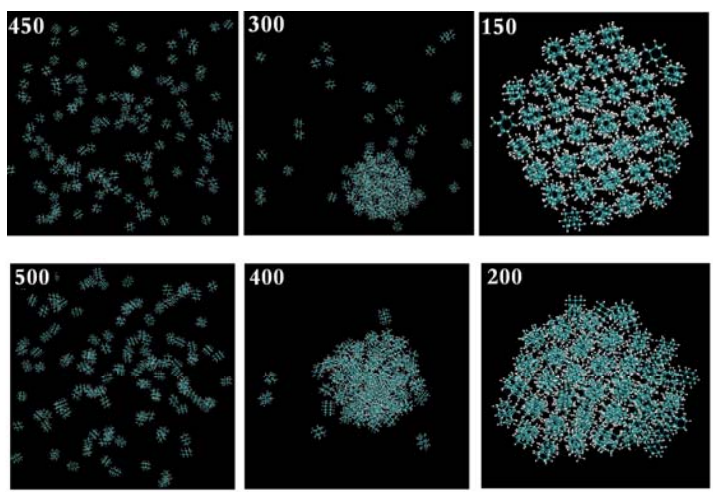

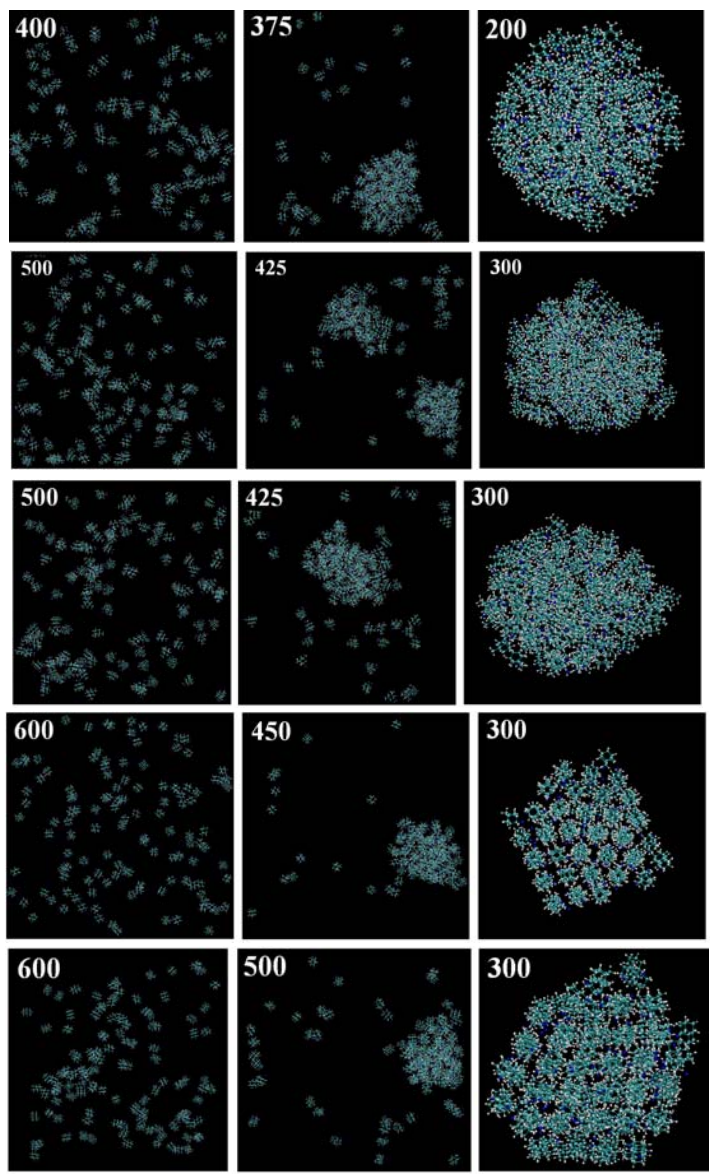

From these snapshots we can directly observe clear phase transitions for each kind of molecule, from the gaseous state to liquid state, and then their aggregation into a highly condensed solid (self-assembled) state.

In order to find the equilibrium configuration of the collection of the 125 molecules at every given temperature we used the simulated-annealing procedure [1, 19]. Every change of 1 K occurred within 10 picoseconds (5,000 time-steps and each time-step was 0.002 ps). With these settings we could observe the self-assembly behaviors in the cooling steps.

It should be mentioned that we applied the cut-off of 2.0 nm for the electrostatics instead of PME, which was used in step d (Figure 2). This was because the simulation box was large enough and system was isolated from other adjacent simulation boxes and it saved us calculation time.

According to Figure 3 adamantane, diamantane, adamantane+Na and diamantane+Na form ordered condensed (crystalline) states. However, amantadine, rimantadine and memantine, while they self-assemble, they do not seem to form clear ordered condensed (crystalline) states. We further produced the hydrogen bonds locations of the same self-assembled snapshots of amantadine, memantine and rimantadine as shown in Figure 4.

**Figure 4**. MD snapshots of (from top to bottom) Amantadine, Rimantadine and Memantine and at 50K hydrogen bonds locations

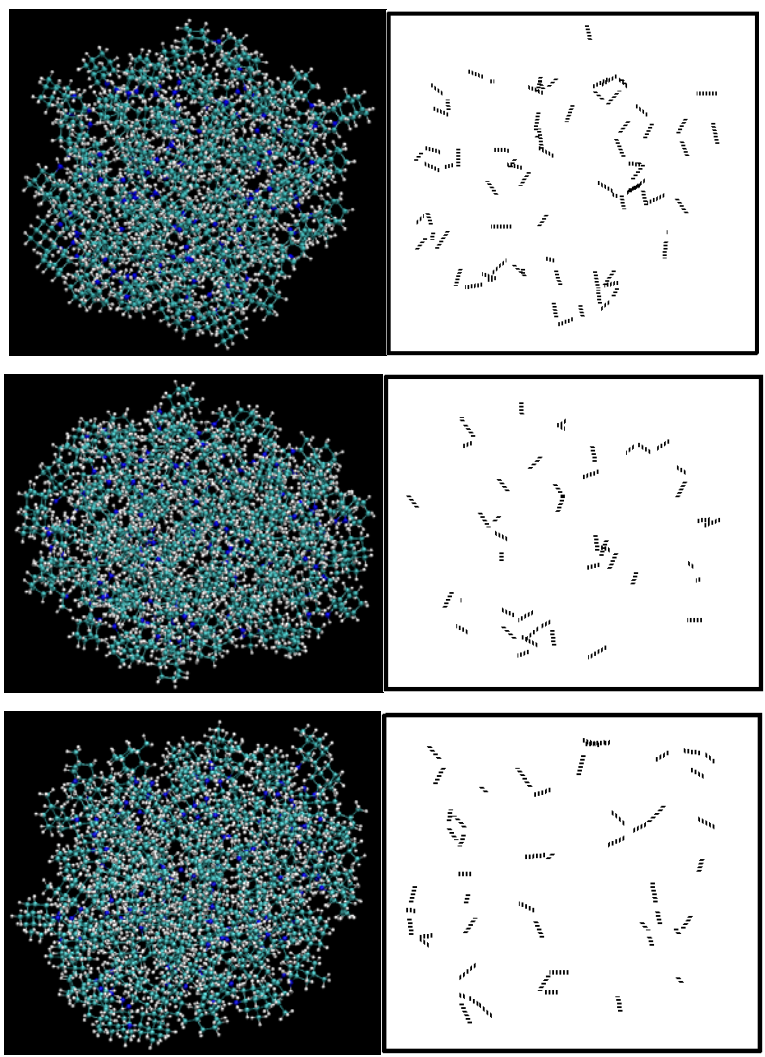

According to this figure we do not observe any ordered format for the location of hydrogen-bonds which is an indication of the non-crystalline self-assembled states of these three kinds of molecules. Obviously the existence of partial hydrogen-bonds between these molecules is the reason for the lack of a clear crystalline state in their self-assemblies.

We also produced the radial distribution functions (RDFs) and structure factors (SFs) of the seven different molecules as presented and discussed below. Study of the RDFs and SFs also reveal further about these phase transition features as discussed below. The RDFs we studied and report here are for the centers of the geometry of molecules.

***3.1. Adamantane and Adamantane+Na***: The main feature of adamantane is that at low temperatures we can observe ordered crystal structures (the top-right image in Figures 3), which matches the previous experimental and theoretical studies [22]. From the RDF and SF graphs at different temperatures (Figure 5) we can observe gas, liquid and solid characteristics.

**Figure 5**. *Radial distribution functions (left) and structure factors (right) of adamantane at 450K, 300K and 150K for gas, liquid and solid states, respectively.*

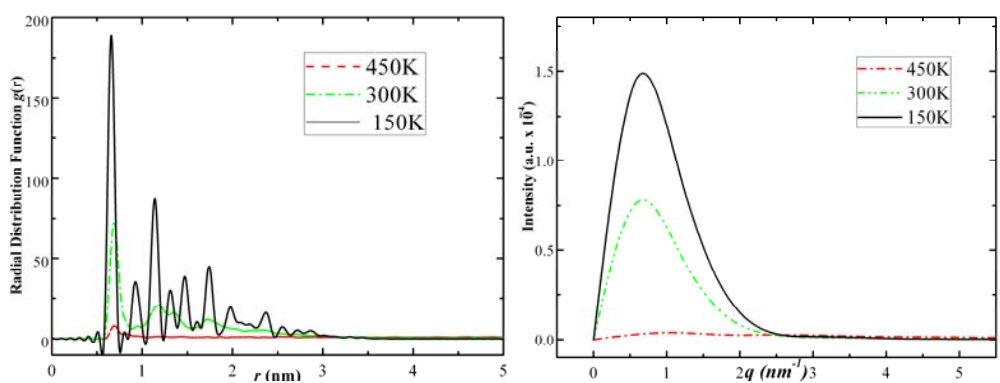

In the RDF figures, taller and sharper peaks can be observed as the temperature decreases and for the SF graphs, the intensity increases as phases transit from gas to liquid and to solid states. We observe similar features in the RDF and SF of adamantane+Na as shown in Figure 6.

**Figure 6**. Radial distribution functions (left) and structure factors (right) of Adamantane+Na at 600K, 450K and 200K for gas, liquid and solid states, respectively.

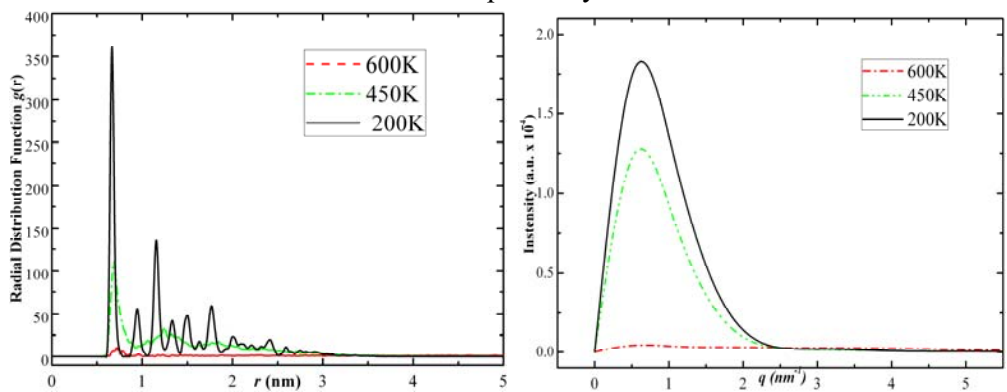

The phase transition temperatures for adamantane+Na are higher than those of adamantine. However, this is attributed to -Na affecting the molecular interactions among adamantanes, i.e. -Na ion in the adamantane+Na structure causes stronger bonding than that of the respective -H ion in adamantane. We may also conclude that the higher phase transition temperatures in adamantane+Na compared to those of adamantane are due to –Na ion, while the geometry structure is determined by the structure of adamantane.

***3.2. Diamantane and Diamantane+Na***: Diamantane molecules can self-assemble to a kind of crystal structure which can be observe from simulation snapshots (Figure 3), however, its self-assembled structure is not as neat as crystal structure of adamantane. In Figure 7 we report the RDF and structure factor of diamantane in various phases.

**Figure 7**. *Radial distribution functions (left) and structure factors (right) of diamantane at 500K, 400K and 300K for gas, liquid and solid states, respectively.*

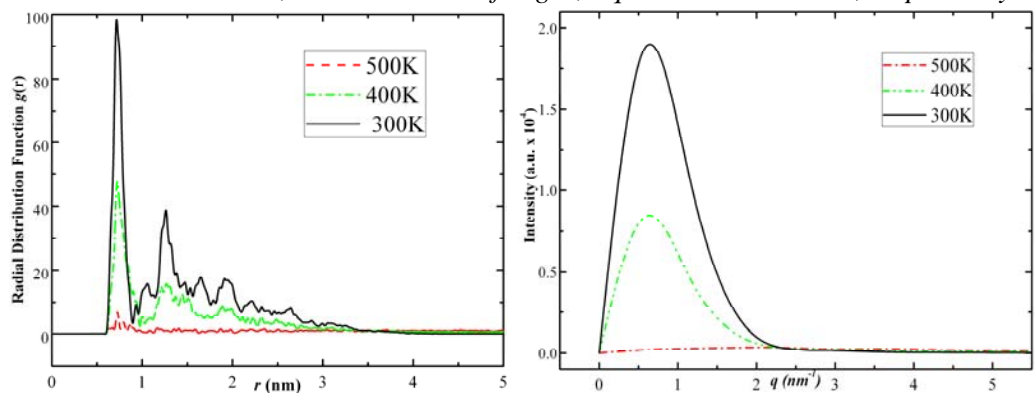

From the RDF and structure factor of diamantane, Figures 7, respectively, we may also conclude the same, i.e. at low temperatures diamantane does not have features of neat crystal structure.

Diamantane+Na seem to have similar relationship to diamantane as adamantane+Na to adamantane, i.e. higher phase transition temperatures due to sodium, and similar crystal structures as diamantane molecules (See related snapshots in Figure 3, RDFs in and SFs in Figure 8).

**Figure 8**. Radial distribution functions (left) and structure factors (right) of Diamantane+Na at 600K, 500K and 400K for gas, liquid and solid states, respectively.

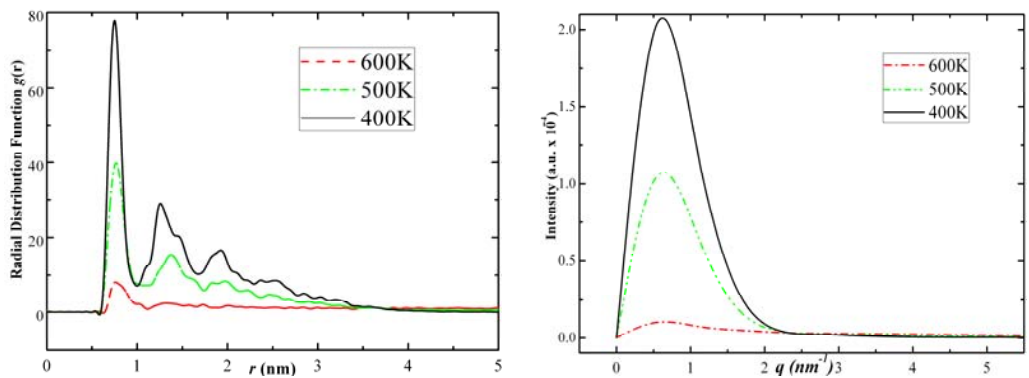

### 3.3. Amantadine, Rimantadine and Memantine:

**3.3. *Amantadine, Rimantadine* and *Memantine*:** The three molecules in this group self-assemble at higher temperatures compared with adamantane, but they don't seem to have well-organized crystalline structures as it is shown in the snapshots in Figure 4. The reasoning for the self-assembly behavior of these three molecules is as follow: Since there are nitrogen atoms in the structure of these molecules, which makes the intermolecular energies much bigger than that of diamondoid, higher temperature should be applied to these systems in order to obtain structures with liquid initial states. When cooling down these molecules, hydrogen-bonds are formed which increases the phase transition temperatures compared to that of adamantane. As it is observed from their self-assembled structures (snapshots in Figure 4) these molecules form certain types of self-assembled structure, however, those structures are not as neat as adamantane and diamantane. There may be two reasons for this: 1. the -NH$_2$ group present in these molecules break some geometrical symmetry of adamantane; 2. we observed that hydrogen bonds are randomly distributed in the bulk structures which makes the entire structures far from an ordered one. In Figures 8-11 we report the RDFs and SFs of these three molecules for the gas, liquid and solid phases.

**Figure 9**. Radial distribution functions (left) and structure factors (right) of amantadine at 450K, 370K and 300K for gas, liquid and solid states, respectively.

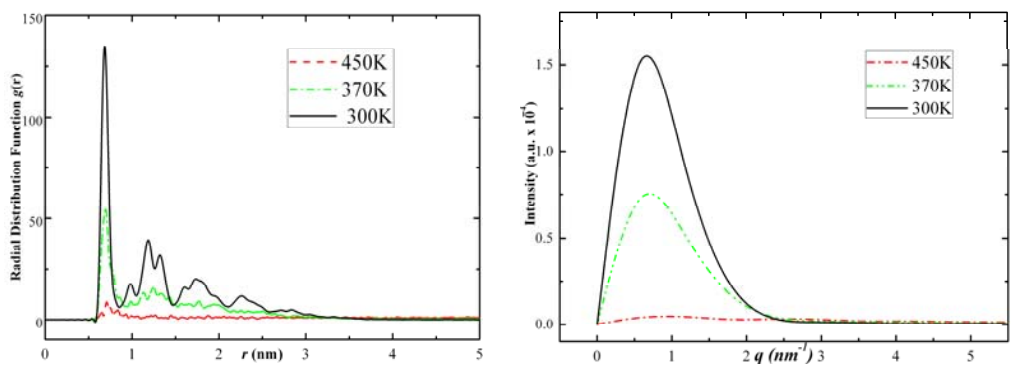

**Figure 10**. Radial distribution functions (left) and structure factors (right) of rimantadine at 500K, 430K and 250K for gas, liquid and solid states, respectively

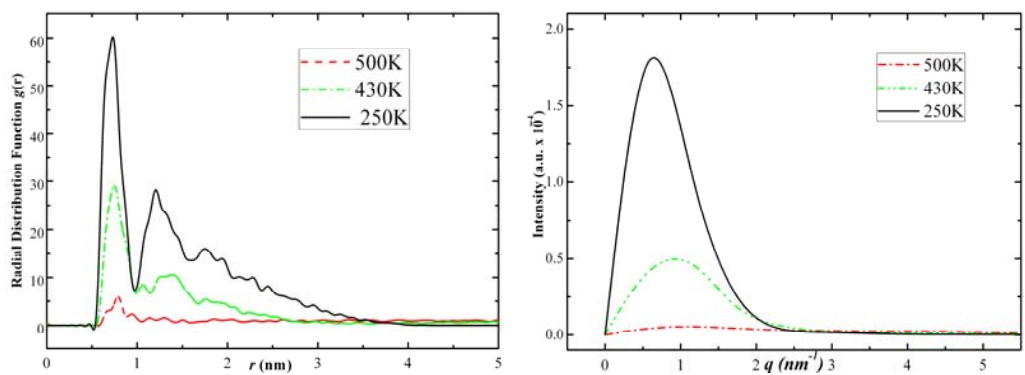

**Figure 11**. Radial distribution functions (left) and structure factors (right) of Memantine at 500K, 430K and 250K for gas, liquid and solid states, respectively.

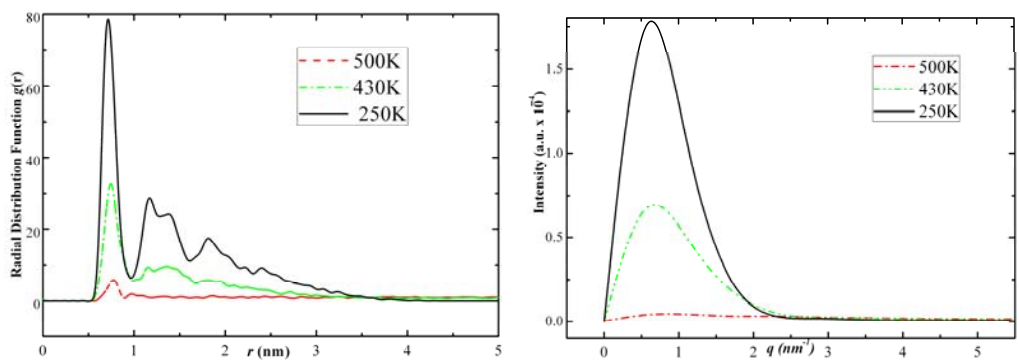

According to Figures 9-11 the solid phase RDFs and SFs analysis of these three molecules show that for most part there is less obvious solid state features, i.e. sharp peaks as in adamantane and adamantane+Na RDFs. The liquid-state RDF graphs for these three molecules also, for most part, have less obvious liquid-state features either.

We also performed studies in the hydrogen-bonds saturation as a function of temperature as well as hydrogen-bond distribution of the three molecules (amantadine, rimantadine and memantine) as shown in Figures 12 and 13.

**Figure 12**. The number of Hydrogen bonds for Amantadine, Memantine and Rimantadine vs. reference temperature. As temperature decreases, the numbers increase, and tend to maximum numbers for the three derivatives.

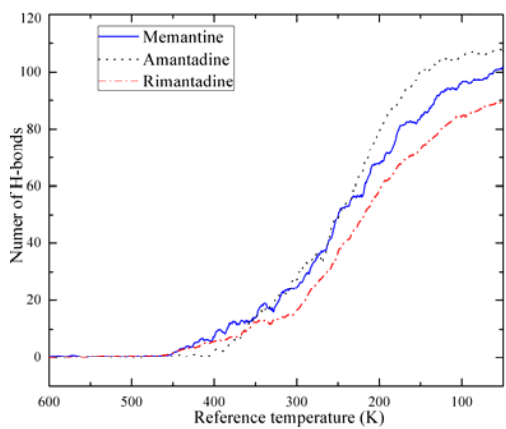

**Figure 13**. Hydrogen-bond distance distribution (Left) and hydrogen-bond angle distribution (Right) at 60K for Amantadine, Rimantadine and Memantine. The most possible hydrogen-bond lengths are around 0.3 nm for the three molecules. The hydrogen-bond angles seem randomly distributed.

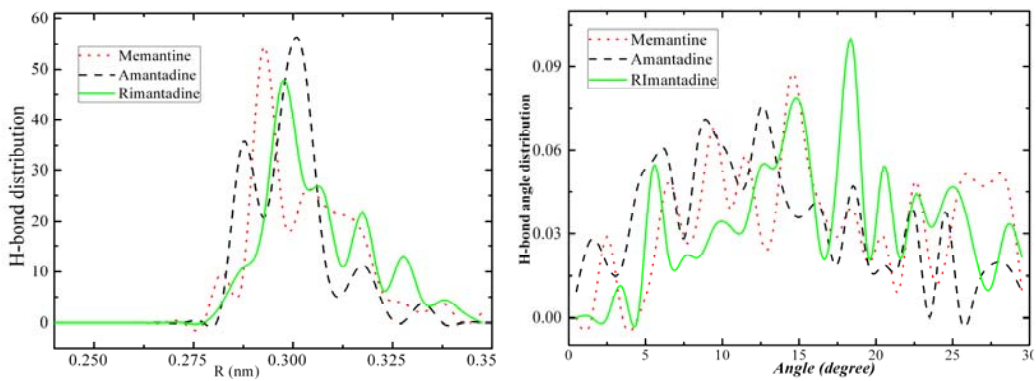

According to Figure 12 in the process of phase transition and self-assembly of these three molecules the number of their hydrogen-bonds saturate at relatively low temperatures (below 100K).

From the hydrogen-bond distance-distribution graph of these three molecules at 60 K (Figure 13-Left), we can find three peaks at about 0.3 nm, which indicate that the structures are close to uniform crystal structures. While the angle distributions (Figure 13-Right) show that the hydrogen-bond angles are rather randomly distributed for all the three molecules. We already mentioned of the hydrogen-bonds of this group at 50 K (Figure 4) which did not show any ordered features. Both Figures 4 and 13 indicate that these structures are not neat crystal structures contrary to adamantane self-assembled state. These three hydrogen-bonded molecules just self-assemble at particular temperatures, but may not form well-organized self-assembled crystal structures. The reason is that the geometry structures of these molecules are not as symmetric as adamantane; and their -NH$_2$ and -CH$_3$ segments, make them unable to pack orderly unlike adamantane.

In Figures 14 and 15 we report the self-assembled (solid-state) radial distribution functions and structure factors, respectively, of the seven molecules with uniform coordinates scales for the purpose of their collective comparison.

**Figure 14**. Radial distribution functions of the seven compounds (from left: Adamantane, Diamantane, Amantadine, Rimantadine, Memantine, Adamantane+Na., Diamantane+Na) in the self-assembled (solid) state.

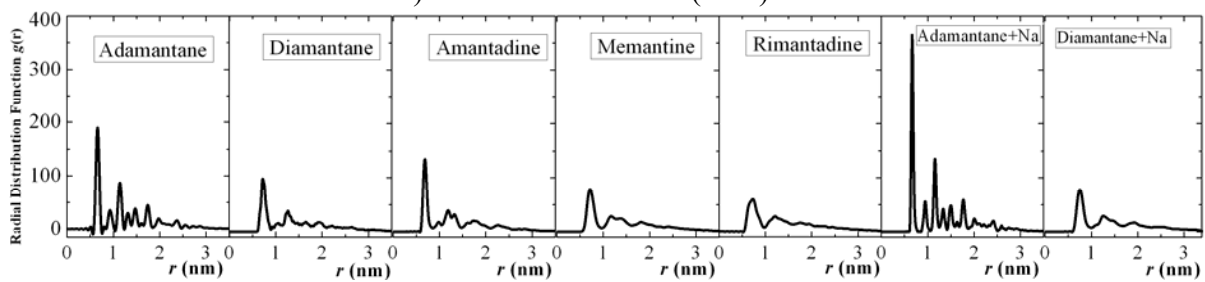

**Figure 15**. Structure factors of the seven compounds (from left: Adamantane, Diamantane, Amantadine, Rimantadine, Memantine, Adamantane+Na., Diamantane+Na) in the self-assembled (solid) state.

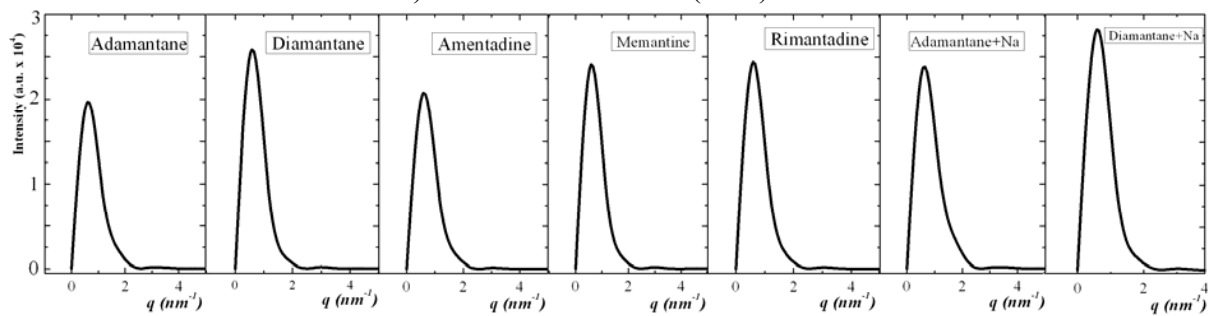

From the latter two figures we can observe that adamantane and adamantane+Na have more sharp peaks than other five molecules, which indicate compact solid state structures. While other five molecules also show solid characteristic peaks but they have less number of such sharp peaks than adamantane and adamantane+Na. These results match the image observation of simulations (Figure 3), i.e. adamantane and adamantane+Na have neater structures in solid state due to their molecular symmetry, which thus prove that self-assembly of those molecules are structure-dependent.

## 4. Conclusions

We have performed a detailed molecular dynamics study of the self-assembly process of seven different diamondoids and derivatives due to temperature variations. From the MD simulation study of the seven molecules, we may conclude the following: 1.The nature of self-assembly in these molecules is a structure-dependent phenomenon. 2. Final self-assembly structures depend on the different bonding types present in the molecular structure of these various molecules. 3. The artificial molecules still hold neat crystal structures. Although -Na ion increases the phase transition temperature, as those the -NH2 ion in group 2, in a large extent the structural features of diamondoids are retained in adamantane-Na and diamantane-Na. The reasons for the latter might be that: A. the -Na ion has less topology effect than does the -NH2 ion. B. there is no hydrogen-bonding in the structures of adamantane-Na and diamantane-Na, therefore they can aggregate to an ordered structures. This feature is very promising, since it allows us to build orderly-shaped NEMS and MEMS.

**Acknowledgment:** We are grateful to the computation facility at University at Illinois at Chicago. All the MD simulation results are visualized by VMD [24].

## 5. References:

[1]. Mansoori, G.A. *Principles of Nanotechnology- Molecular-Based Study of Condensed Matter in Small Systems*; World Scientific Pub. Co., Hackensack, NJ, **2005**.

[2]. Mansoori, G.A. Diamondoid Molecules. *Advances in Chemical Physics*, **2007**; 136, 207-258.

[3]. Zhang, G.P.; George, T.F.; Assoufid, L.; Mansoori, G.A. First-principles simulation of the interaction between adamantane and an atomic-force microscope tip. *Phys. Rev. B*, **2007**, 75, 035413.

[4]. Ramezani, H.; Mansoori, G.A.; Saberi, M.R. Diamondoids-DNA Nanoarchitecture: From Nanomodules Design to Self-Assembly *J. Comput'l & Theor'l Nanoscience* **2007**, 4(1), 96-10.

[5]. Mansoori, G.A.; George, T.F.; Zhang, G.P.; Assoufid L, Eds. Molecular Building Blocks for Nanotechnology: From Diamondoids to Nanoscale Materials and Applications, *Topics in Applied Physics* Springer, New York, **2007**,109, 44-71.

[6]. Ramezani, H; Mansoori, G.A. Diamondoids as Molecular Building Blocks for Nanotechnology. *Topics in Applied Physics*, Springer, New York, **2007**, Topics in Applied Physics **109**, 44-71.

[7]. Ramezani, H.; Saberi, M.R.; Mansoori, G.A. Diamondoids and DNA Nanotechnologies. *Int'l J. Nanoscience & Nanotechnoogy* (*IJNN*), **2007**; 3(1): 21-35.

[8]. Xue, Y.; Mansoori, G.A. Quantum Conductance and Electronic Properties of Lower Diamondoids and their Derivatives. *International Journal of Nanoscience*, **2008**, 7(1), 63 – 72.

[9]. Marsusi, F.; Mirabbaszadeh, K.; Mansoori, G.A. Opto-electronic Properties of Adamantane and Hydrogen-Terminated Sila- and Germa-Adamantane: A Comparative Study. *Physica E* **2009** 41, 1151-1156.

[10]. Mansoori, G.A.; George, T.F.; Zhang, G.P.; Assoufid, L. Structure and Opto-Electronic Behavior of Diamondoids, with Applications as MEMs and at the Nanoscale Level, *Progress in Nanotechnology Research*, Nova Pub's. **2009**, Chapter 1, 1-19.

[11]. Kohn, W.; Sham, L. Self-Consistent Equations Including Exchange and Correlation Effects. *Physical Review* **1965**, 140, A1133-A1138.

[12]. Lee, C.; Yang, W.; Parr, R.G. Development of the Colle-Salvetti correlation-energy formula into a functional of the electron density. *Phys. Rev. B* **1988**, *B* 37, 785-789.

[13]. Woon, D.E.; Dunning T.H. Jr. Gaussian basis sets for use in correlated molecular calculations. III. The atoms aluminum through argon. *J. Chem. Phys.* **1993**, 98, 1358.

[14]. Curtiss, L.A.; McGrath, M.P.; Blaudeau, J-P.; Davis, N.E.; Binning, R.C. Jr.; Radom, L. Extension of Gaussian-2 theory to molecules containing third-row atoms Ga-Kr. *The Journal of Chemical Physics* **1995**, 103**,** 6104–6113.

[15]. Frisch, M.J.; Trucks, G.W.; Schlegel, H.B.; *et al*. Gaussian 03, Revision D.01, Gaussian, Inc., Wallingford CT, 2004.

[16]. Carpenter, J.E.; Weinhold, F.J. Analysis of the geometry of the hydroxyl radical by the ldquo. *Mol. Struct. (Theochem)* **1988**, 169, 41-62.

[17]. Jorgensen, W.L.; Maxwell, D.S.; Tirado-Rives, J. Development and Testing of the OPLS All-Atom Force Field on Conformational Energetics and Properties of Organic Liquids. *J. Am. Chem. Soc.* **1996**, 118, 11225–11236.

[18]. Damm, W.; Frontera, A.; Tirado-Rives, J.; Jorgensen, W.L. OPLS All-Atom Force Field for Carbohydrates. *Journal of Computational Chemistry* **1997**, 18, 16, 1955-1970.

[19]. Lindahl, E.; Hess, B.; van der Spoel, D. GROMACS 3.0: A package for molecular simulation and trajectory analysis. *J. Mol. Mod.* **2001, 7,** 306-317.

[20]. van der Spoel, D.; Lindahl, E.; Hess, B.; Groenhof, G.; Mark, A.E.; Berendsen, H.J.C. GROMACS: Fast, Flexible and Free. *J. Comp. Chem.* **2005**, 26, 16, 1701-1718.

[21]. Hess, B.; Kutzner, C.; van der Spoel, D.; Lindahl, E GROMACS 4: Algorithms for highly efficient, load-balanced, and scalable molecular simulation. *J. Chem. Theory Comput.* **2008**, 4, 435-447.

[22]. Greig, D.W.; Pawley, G.S. Molecular dynamics simulations of the order-disorder phase transition in adamantane. *Molec. Phys.* **1996**, 89, 3,

677-689.

[23]. Ciccotti, G.; Ferrario, M.; Memeo, E.; Meyer, M. Structural Transition on Cooling of Plastic Adamantane: A Molecular-Dynamics Study. *Phys. Rev. Lett.* **1987**, 59, 2574-2577.

[24]. Humphrey, W.; Dalke, A.; Schulten, K. VMD-Visual Molecular Dynamics. *J. Molec. Graphics* **1996**, 14, 33-38.